# The one-way function based on computational uncertainty principle

P.F. Wang[1], J.P. Li

State Key Laboratory of Numerical Modeling for Atmospheric Sciences & Geophysical Fluid Dynamics (LASG), Institute of Atmospheric Physics, Chinese Academy of Sciences, Beijing, 100029，China

**Abstract** This paper presents how to make use of the advantage of round-off error effect in some research areas. The float-point operation complies with the reproduce theorem without the external random perturbation. The computation uncertainty principle and the high nonlinear of chaotic system guarantee the numerical error is random and departure from the analytical result. Combining these two properties we can produce unilateral one-way function and provide a case of utilizing this function to construct encryption algorithm. The multiple-precision (MP) library is used to analyze nonlinear dynamics systems and achieve the code. As an example, we provide a scheme of encrypting a plaintext by employing the one-way function with Lorenz system. Since the numerical solution used in this scheme is beyond the maximum effective computation time (MECT) and it cannot satisfy the requirements of return-map analysis and phase space reconstruction, it can block some existing attacks.

**Keywords** computational uncertainty principle, MECT, one-way function, encryption

## 1 Introduction

The late 1990s saw the boom of using the dynamical properties of chaotic systems to encrypt a message, or a 'chaotic encryption approach' [1,2]. The stream cipher scheme based on the Lorenz equation was then brought forth, and its security caused another round of discussion at the turn of the 21$^{st}$ century[3–5]. It is understood that the dynamical degradation will occur when chaotic system is used in digital cipher, and this degradation will threaten the security of digital chaotic cipher[6]. The return-map attack method was first proposed by P´erez and Cerdeira[7] to attack chaotic switching and chaotic masking schemes based on the Lorenz system and it was restudied to break other chaotic based cipher system. Vaidya et al[8] found a quick way to identify the superkey (the three parameters of the equations) of the Lorenz equation, and claimed the message can therefore be extracted easily. Li et al[9] also pointed out that some of the previous chaotic encryption schemes are not secure when they are computerized with finite computing precision.

---

[1] corresponding author: Pengfei Wang, LASG, P.O.Box 9804, Beijing 100029, P.R.China; Email:wpf@mail.iap.ac.cn; Fax:86-010-82995172

In this study we combined our knowledge of nonlinear equation round-off error effect with chaotic cryptography and provided the concept of one-way code generator that can satisfy the requirements of cryptography and therefore can produce secure encryption of data.

## 2 The Computational Uncertainty Principle

The study of round-off error in numerical computation can go all the way back to the time before the modem computer was invented. It was discussed by astronomers[10,11] then the pioneering important work on the analysis of numerical error with round-off error can be found in the Neumann[12] and Tuing's[13] paper soon after the first computer was invented. Later Wilkinson[14] and Henrici[15,16] published books on the round-off error in algebraic and difference process. The more comprehensive introduction of round-off study can be found in the book by Higham[17] and the reference cited therein. Most discussion of round-off error is about how they cause the shortage of stability and convergence etc, and the behavior is still far beyond analytical analysis. But in the following work, we will present that the difficulty to determine the result of round-off error may have the surprising advantages in some research areas.

The studies of the round-off error in nonlinear dynamical system integration[18,19] in the late 20$^{th}$ century, showed that the Computational Uncertainty Principle (CUP) exists with certain precision (single and double precisions) in the computation. We[20] extended this research result by applying multiple-precision (MP) method to do accurate experiments within the Lorenz system, and obtained new computational characteristics with and without chaos condition. We found that when dynamical system is in chaos status, the effective computation time (ECT) exists in the numerical solution of nonlinear equation. With certain computing precision, the numerical result is indeterminate and sensitive to step-size when time t is greater than the ECT and the different effective computation precision (ECP) exists in the numerical solution of nonlinear equation corresponding to different time. We can obtain the numerical solution close to the real value only if the precision we use in the method is equal or greater than the ECP, The numerical result is determinate when the initial value, method, and step-size are constant while the precision is infinite or finite, but the result may not be the real value.

The classical Lorenz equations introduced by Lorenz[21] are as follows:

$$\begin{cases} \dfrac{dx}{dt} = -\sigma x + \sigma y \\ \dfrac{dy}{dt} = \gamma x - y - xz \\ \dfrac{dz}{dt} = xy - \beta z \end{cases} \quad (1)$$

where $\sigma, \gamma, \beta$ are nondimensional constants, and t is nondimensional time. Here, we consider Eq.1 with chaos, when $\gamma =28.0$, $\sigma =10.0$, and $\beta =8/3$. The initial value is set as (5,5,10).

Before we start to discus one way function generated by float-point process, we shall put forward some mathematical and computational principle.

**Theorem 1.** (from Li et. al 2001)

When using difference method to solve the ordinary different equations (1), the total numerical error (include round-off error) satisfies the formula from statistical view:

$$\left\| E(t,h) \right\| = \left\| E_t(t,h) + E_r(t,h) \right\| \simeq C(t)\left( Ch^m + \dfrac{\sigma}{\tilde{C}\sqrt{h}} \right) \quad (2)$$

where $E_t(t,h)$ is the discretization error, $E_r(t,h)$ is the round-off error, $C(t)$ is a time function, $C$ is constant depend on method, $\sigma$ is constant depend on precision, $\tilde{C}$ is constant depend on ODEs, $m$ is the method order and $h$ the time step-size.

The minimize absolute numerical error exists when the numerical difference methods are used to solve ODES in a finite precision computer when h satisfies

$$h = \left(\frac{\sigma}{2mC\tilde{C}}\right)^{1/(m+0.5)} \tag{3}$$

More details and the proof of (2) and (3) can be found in Li et al. (2001).

**Theorem 2[19].**

With the same numerical method of order $m$ in two precision $p1$ and $p2$, then

$$T_2 - T_1 = \hat{C} \cdot m \cdot \ln\left(\frac{H_1}{H_2}\right)^m = \hat{C} \cdot m \cdot \ln 2^{\frac{p_2-p_1}{m+0.5} \cdot m} \tag{4}$$

where $T_1$ and $T_2$ are the MECT corresponds to precision $p1$ and $p2$. $H_1$ and $H_2$ are the OS corresponds to precision $p1$ and $p2$, $\tilde{C}$ is constant depends on ODES.

This formula can be convenient used to estimate MECT or precision.

**Lemma 1.**

In a certain float point system, the basic float-point operation $\oplus, \ominus, \otimes, \oslash$ (corresponds to the +,-,*,/ which is the mathematics operation within real numbers) in computer can be regard as $op$, and the result of two variable with basic operation $c = a(op)b$ is deterministic.

This can be proved by the uniform expression of real numbers and the uniform round-off method applied within the same float-point operation computation system.

**Lemma 2.**

The complex algebraic process can be broken down into a set of basic operation.

$$c = a_1(op)a_2(op)a_3\cdots(op)a_n$$

From lemma 1 we know that each of the single basic operation result is determinate. So the final result of the algebraic process is also deterministic.

**Theorem 3.** (Special reproduce theorem of float-point computation)

The numerical difference method to solve ODEs with the initial conditions will get a deterministic result when there is no external pertubation.

This can be deduced by lemma 2 since the difference method can be looked as a serial of algebraic computation process.

This theorem guarantee that we will get the same result by using the same program and the same parameters in the same platforml.

**Theorem 4.**

In despite of the numerical difference method can get deterministic result but the result may be different from the mathematical result because the numerical error exist in it.

The previous work indicated that the MECT of Eq.1 with single precision is about 17, and the MECT of double precision is about 35 (the 4-th order RK method). We can us formula (4) to measure the MECT time of 256 bits precision.

$$
\begin{aligned}
35 - 17 &= \hat{C} \cdot 4 \cdot \ln 2^{\frac{53-24}{4+0.5} \cdot 4} \\
T_{256} - 17 &= \hat{C} \cdot 4 \cdot \ln 2^{\frac{256-24}{4+0.5} \cdot 4}
\end{aligned}
\quad (5)
$$

The analytical solution of (5) is $T_{256} \simeq 162$, and it is less than 200. The MP experiment showed that with 256-bit precision the MECT is the same as above. Thus when $t >= 200$, we can't obtain the right solution close to the real value. Moreover, the numerical result is different when time-steps size h varies. The reason is that with the selected precision, a long enough time t can lead to unpredictable numerical results which only depend on the coefficients $\sigma, \gamma, \beta$, the initial value $x_0, y_0, z_0$ and step-size. All previous return-map and phase space reconstruction analyses have a common premise that the value series is the true value from the original equation, so when we choose the numerical result beyond the MECT, the value is not the true value and therefore it cannot fit this premise. For example the study of Li et al[22] showed that when the numerical solution is beyond MECT, the fractal dimension estimating result is sensitive to the time intervals, and the one-dimension time series result is different from the three-dimension time series.

**3 One-way mapping based on CUP**

Since the variables $\sigma, \gamma, \beta, x_0, y_0, z_0, h, p, t$ can determine the numerical results, where $h$ is step-size, $p$ is the computation precision, we can define the algorithm as:

$$(x, y, z) = L(\sigma, \gamma, \beta, h, p, t, x_0, y_0, z_0),$$

where $(x, y, z)$ can be regard as a vector $A$, and $L$ can be any numerical integrating method. In this paper the 4$^{th}$ Runge-Kutta (RK) method is used, $\sigma, \gamma, \beta, x_0, y_0, z_0, h, p, t$ are input variable, and $A$ is output.

From the algorithm:

$$A = L(\sigma, \gamma, \beta, h, p, t, x_0, y_0, z_0),$$

in the integration of $A$, the process can be devided into n steps, and in each step $(x_i, y_i, z_i) = L(\sigma, \gamma, \beta, h, p, t, x_{i-1}, y_{i-1}, z_{i-1})$. We assume $n = \left[\frac{t}{h}\right]$ and if $t$ can't be evenly divided by h, the last step $h' = t - n*h$. Since both $L$ and $\sigma, \gamma, \beta, x_0, y_0, z_0, h, p, t$ are determinate, from the theorem 1 we know that the $A$ is therefore also determinate.

The choice of $\sigma, \gamma, \beta, x_0, y_0, z_0, h, p, t$ is infinite while the output vector $A$ is finite, therefore $L$ is a multi-to-one mapping and collision maybe occur. We can choose 256-bit or higher $A$, and then the probability of collision is very small in practice. The inverse function of $L$ does not exist and therefore even if $A$ is known, the unique variable sets $\sigma, \gamma, \beta, x_0, y_0, z_0, h, p, t$ can still not be obtained.

The algorithm $A = L(\sigma, \gamma, \beta, h, p, t, x_0, y_0, z_0)$ is a one-way or irreversible mapping. We can then define a hash function based on CUP and furthermore a CUP-based chaotic stream cipher. We call the encryption method based on this function the "Lorenz Code".

One of the cipher code generating procedures is convert message to input variables. When we convert a certain string to $\sigma, \gamma, \beta, x_0, y_0, z_0, h, p, t$ this convert routine is one to one correspondence. In order to keep the chaotic behavior of the Lorenz system, we must maintain $\gamma > 28.0$. If the output message is required to be 256 bits, the precision $p$ must be large or equal to 256 bits. The time $t$ should be larger than 200. Considering the encryption speed, if $p$ is chosen as 256 bits, it's more convenient to have $t$ less than 1000. The other variable should be converted within a certain range, neither too big nor too small, and $h$ is at about magnitude 0.001.

As an example, we assign the following nine basic parameters with different values: $\gamma = 28.0$, $\sigma = 10.0$, $\beta = 8/3$, $(x_0, y_0, z_0) = (5, 5, 10)$, $h = 0.01$, $p = 256$, $t = 200$. We keep $p = 256$, and set the input message as 8 ASCII code: $m_1 m_2 m_3 m_4 m_5 m_6 m_7 m_8$. The one variable such as $x$ of the three output results is used as encrypted value to resist the return-map attack and phase space reconstruction.

Because $m_1$ ranges from 0 to 255, we can use $m_1 / 1000$ as the perturbation of $\gamma$, and set $\gamma' = \gamma + m_1 / 1000$. The other parameters are converted as follows: $\sigma' = \sigma + m_2 / 1000$, $\beta' = \beta + m_3 / 1000$, $x' = x + m_4 / 1000$, $y' = y + m_5 / 1000$, $z' = z + m_6 / 1000$, $h' = h + m_7 / 1000$, $t' = t + m_8$. The numerical integration result is corresponding to the message $m_1 m_2 m_3 m_4 m_5 m_6 m_7 m_8$. The output message is 256 bits. It can be converted to a 32-byte ASCII string, after the radix point is removed.

When the above approach is used as hash function, $\sigma, \gamma, \beta, x_0, y_0, z_0, h, t$ are parameters and $m_1m_2m_3m_4m_5m_6m_7m_8$ is input message. We can use this hash function to compute the hash string of any 8 bytes messages. No collision is found with about 8 million different input messages, which proves that the hash function works well. We also did random number test of the output messages by applying diehard[23] software, and the result was acceptable.

**4 Chaotic stream cipher method based on Lorenz Code**

When this approach is used as chaotic stream cipher generator, $m_1m_2m_3m_4m_5m_6m_7m_8$ is secure key and $\sigma, \gamma, \beta, x_0, y_0, z_0, h, t$ are parameters. We can encrypt the plaintext into ciphered messages. For example, the plaintext is known as: $M_1M_2M_3...M_k$. It can be divided into some 32 bytes string groups and the zero is added if the last group has less than 32 bytes. For the first group $M_1M_2M_3...M_{32}$, we use $m_1m_2m_3m_4m_5m_6m_7m_8$ to generate a secure string $K_1K_2K_3...K_{32}$ and use it to operate with $M_1M_2M_3...M_{32}$ to generate $C_1C_2C_3...C_{32}$. Then we send the 32 bytes string $C_1C_2C_3...C_{32}$ as the first ciphered message. We use $M_1M_2M_3...M_{32}$ and $C_1C_2C_3...C_{32}$ as the feedback message to operate with $m_1m_2m_3m_4m_5m_6m_7m_8$ and get another string $n_1n_2n_3n_4n_5n_6n_7n_8$ to encrypt the second group of plaintext. We repeat this encryption operation until all plaintext groups are done.

When the ciphertext is received, the receiver can divide the message into some 32 bytes message groups and use the encryption method $L$ to encrypt the secure key $m_1m_2m_3m_4m_5m_6m_7m_8$ to obtain $K_1K_2K_3...K_{32}$, which can be operated with $C_1C_2C_3...C_{32}$ to regenerate the plaintext $M_1M_2M_3...M_k$. Then the receiver can use the $M_1M_2M_3...M_{32}$ and $C_1C_2C_3...C_{32}$ as the feedback message to operate with $m_1m_2m_3m_4m_5m_6m_7m_8$ and get another string $n_1n_2n_3n_4n_5n_6n_7n_8$ to decrypt the second ciphertext. The receiver repeats the above procedures, and the whole ciphertext can be decrypted.

**5 Conclusion**

The studies of nonlinear dynamical system, especially the round-off error effect in computation, have influenced many scientific fields. In this research, we use it in encryption study and propose the encrypt scheme based on CUP-MP, or a MP based Lorenz encryption scheme employing the CUP theory.

The reproduced theorem in the float-point system can guarantee the deterministic numerical result. The computation uncertainty principle and high nonlinear chaotic of Lorenz system guarantee that the numerical error is random and departure from the analytical result. Combining these two facts together we get the numerical one-way function.

We also present the method to use numerical one-way function to construct 'The Lorenz Code' encrypt method, which is feasible, secure and its output bits are easy to control. It is easy to produce a key of 256 bits or even higher. Since the algorithm is irreversible the common enumerate attacks cannot easily break it.

The connection between round-off error of chaotic nonlinear system computation and cryptography is established, and the idea and scheme can be easily adapted to any other chaotic systems. Origination from this new idea, thousands of new encryption-with-chaos systems could be produced and therefore it is not necessary to make special secure analysis to each system.

**Acknowledgements**

We would like to thank Dr. Shujun Li for his valuable suggestions on the secure test.